\documentclass[10pt,reqno]{amsart}
\usepackage{amsmath}
\usepackage{graphicx}
\usepackage{epsfig}
\usepackage{mathrsfs}
\theoremstyle{plain} 

\newtheorem{thm}{Theorem}

\theoremstyle{definition}
\newtheorem{defn}{Definition}

\theoremstyle{remark}

\numberwithin{equation}{section}

\DeclareMathOperator{\prob}{Pr} \DeclareMathOperator{\Esp}{E}

\DeclareMathSymbol{\R}{\mathalpha}{AMSb}{"52}
\DeclareMathSymbol{\C}{\mathalpha}{AMSb}{"43}

\newcommand{\pd}{\partial}

\newcommand{\ra}{\rightarrow}

\begin{document}

\title[Classification of barrier options]{
Classification of barrier options}
\author[J.C. Ndogmo]{J.C. Ndogmo}

\address{P O Box 2446\\
Bellville 7535\\
South Africa.}
\email{ndogmoj@yahoo.com}

\begin{abstract}
For a given level of accuracy in option prices, the paper considers
the problem of deciding when exactly, as one or more of the pricing
parameters change, a barrier option degenerates into a simpler type
of option. This problem is meaningful in the real world where option
prices are always determined within a certain level of accuracy. The
problem is reduced to finding certain critical values of the initial
stock price, and this is achieved through a probability-based
approach.
\end{abstract}


\subjclass[2000]{91b24, 65c30}
%
\maketitle

\section{Introduction}
\label{s:intro}%
One of the topics that financial mathematics has been concerned with
over the last decades is that of option pricing, and barrier option
pricing features prominently among these topics, in particular
because barriers can be added to any existing option. In this
regard, Merton \cite{merton} obtained in $1973$ the first pricing
formula for a European down-and-out call option by solving
explicitly the associated differential equation with the appropriate
boundary conditions. Latter on in 1985, Cox and Rubistein~
\cite{cox} obtained a similar formula for the up-and-out barrier.
Using a probabilistic approach based on certain theoretical results
derived earlier on by Levy~ \cite{levy} and by Anderson~
\cite{ander}, Kunitomo and Ikeda~ \cite{kuni} obtained more general
pricing formulas for European double barrier options with  curved
barriers and for a variety of path-dependent options and corporate
securities, some of which had already been independently obtained
for restricted cases in \cite{coxb,inger, connell}.\par
 Attempts to extend these pricing formulas to the more complex case of
American-style options have been undertaken in papers by Broadie and
Detemple~ \cite{broadie}, Gao {\em et al}~ \cite{gao}, and Haug~
\cite{haug2}. All of these extensions to American-style options have
however only lead to close-form approximations. In parallel with the
determination of these pricing formulas, numerical methods have been
used for pricing barrier options, especially in those cases where
analytical pricing solutions remain unavailable, such as for
discrete barrier options \cite{MC, TT, MCh, fusai}, or for
American-style options that, in some cases, comprise transaction
costs or stochastic volatilities \cite{barraq, gao, clarke,
ikonen}.\par

An important problem about barrier options that doesn't seem to have
been considered is that of deciding when exactly, for a given level
of accuracy in option prices, a double barrier option degenerates
into a much elementary type of barrier option, as one or more of the
pricing parameters change with time. Although this problem is
meaningless in terms of the Brownian motion of the of the underlying
asset price, it becomes fully meaningful in the real word where all
prices are known only to within a finite number of precision digits.
It is also a practical problem in the sense that it can lead to a
more efficient pricing of various types of barrier options, and to a
better insight into the dynamics of options, and more specifically
barrier options.\par

In this paper, we consider the problem of determining when exactly,
as the initial stock price or other pricing parameters evolve, and
for a fixed level of accuracy in option prices, a barrier option
with a single or a double barrier degenerates into a simpler type of
option, which may be, for instance in the case of a double barrier
option, a down-and-out or an up-and-out barrier option, or just a
vanilla. We consider in this analysis more general types of barriers
with time-dependent absorbing boundaries. The problem is reduced to
finding certain critical values of the initial asset price, and we
use a probabilistic approach to derive close-form approximate
solutions for these values. Numerical experiments based on
analytical option pricing results such as those from \cite{kuni} are
then implemented, to check the validity of the formulas
obtained.\par
%

In the next section, we derive formulas for the classifying critical
asset prices which in  Section $3,$ are applied to the problem of
classification of barrier options. We present the numerical test of
these formulas in  Section $4,$ and give some concluding remarks in
Section~ \ref{s:conclusion}. Due to the usual parity considerations,
we focus our analysis solely on knock-out barrier options.

\section{Classifying critical asset prices}\label{s:critical}

\begin{figure}[ht]
\begin{center}
\caption{\label{fg:path1} \protect \footnotesize Sample asset price
path together with lower and upper barriers. \vspace{2mm}}
\includegraphics[width= \textwidth]{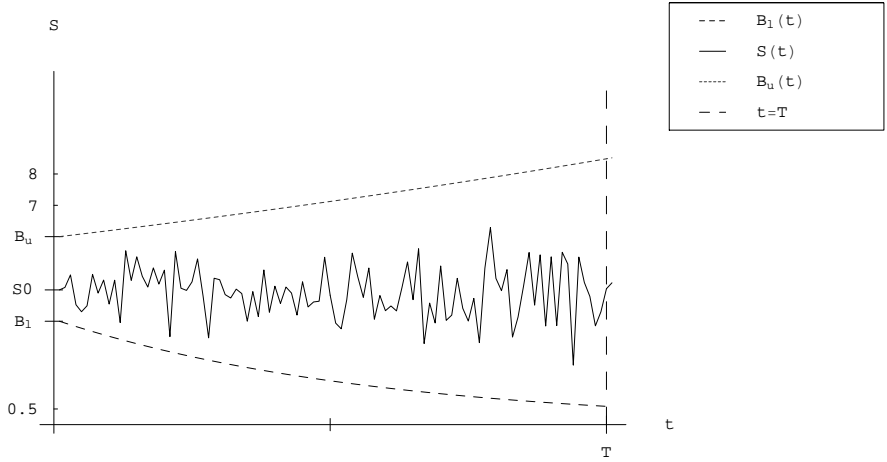}
\end{center}
\end{figure}

We make the usual assumption that the stock price $S_t$ follows a
geometric Brownian motion, or equivalently, that it is lognormally
distributed. In this case, the stochastic differential equation
satisfied by $S_t= S(t)$ is given in terms of the standard Wienner
process $W$ by
\begin{equation} \label{eq:gbm} d S_t= S_t \mu dt + S_t \sigma d W, \end{equation}
where $\mu$ and $\sigma$ are the constant drift and volatility
parameters, respectively. Consider two continuous curves $S= B_l(t)$
and $S=B_u(t)$ over the time interval $I=[0,T],$ such that $B_l <
S_0 < B_u,$ where $B_l=B_l(0)$,  $B_u= B_u(0),$  $S_0=S(0)$ and $T$
is a given time. Assume furthermore that the two curves are
non-overlapping, so that we have $B_l(t)< B_u (t)$ for all $t \in
I.$  A sample path of stock price $S_t$ together with the two
adjacent curves are depicted in Figure~ \ref{fg:path1}.\par

%
%
%
 Denote by $\theta= \theta(f)$ the accuracy
in a given option price $f.$ For simplicity, we may write
$\theta(f)$ in the form $\theta(f)= 10^{- \mathsf{m}},$ where
$\mathsf{m}$ is the number of significant digits to the right of the
decimal point in $f.$ Suppose now that $f$ is the price of a double
knock-out option contingent on $S_t,$ with curved barriers $B_u(t)$
and $B_l(t).$ For any option under consideration, unless otherwise
stated we denote by $S_0$ the initial price of the underlying asset,
by $K$ its strike price and by $T$ its expiry date. Let $\mathsf{v}$
be a variable symbol that may take on the value $u$ to mean "upper"
or $l$ to mean "lower". Let $E_\mathsf{v}$ be the event that the
$\mathsf{v}$ barrier is breached before the other barrier by time
$T,$ and let ${\rm R}_\mathsf{v}$ be the corresponding rebate paid
at the expiry date of the option. If we denote the risk-free
interest rate by $r,$ by $\prob$ the probability operator, and by
$\Esp$ the risk neutral expectation operator, then the price $f$ of
the double knock-out option is given by
\begin{equation}\label{eq:gndbl}
f= e^{-r T}\Esp \left[ {\rm R}_u \prob (E_u)+ {\rm R}_l \prob(E_l) +
(1-\prob(E_u)-\prob(E_l)) \prob (S_T > K) (S_T -K)\right]
\end{equation}

   In terms of the geometric Brownian motion of the underlying, the
probability that the stock price breaches a barrier before any given
time $T$ is never zero. Note that the operator $\Esp$ is additive
with respect to each term in its argument in \eqref{eq:gndbl}. By
the same properties of the geometric Brownian motion, by choosing
$S_0$ sufficiently far away from an initial barrier value
$B_\mathsf{v},$ we can make $\prob (E_\mathsf{v})$ so small that if
we let $\Delta f$ be the difference between the corresponding value
of $f$ and the value of $f$ obtained by setting $\prob
(E_\mathsf{v})=0,$ then $\Delta f < 5\times 10^{-1} \theta (f).$
This in terms of the accuracy of $f$ means that $\Delta (f)=0,$ and
thus the barrier $B_\mathsf{v}(t)$ has therefore no effect on the
value of $f$, and the double knock-out option degenerates into a
simpler type of option, which may be a down-and-out barrier option
if $\mathsf{v}=u,$ or an up-and-out barrier option if
$\mathsf{v}=l.$ In the real world markets where all prices are
determined only in terms of a finite number of precision digits, we
may thus assume that the probability of a stock price breaching a
barrier before a given time $T$ can take on the value zero for
appropriately chosen values of the initial price $S_0$.\par

It is therefore meaningful to consider the following two
complementary problems, for a given accuracy $\theta$ of option
prices.
\begin{description}
\itemsep = 1mm
\topsep = 2.5mm

\item[{\small Problem 1}] Find the minimum value $S_{ml}$ of $S_0$ which ensures that
the stock price will not breach the lower curve before time $T.$
\item[{\small Problem 2}] Find the maximum value $S_{mu}$ of $S_0$ which ensures
that the stock price will not breach the upper curve before time
$T.$
\end{description}

\begin{defn}
We say a curve $S=g(t)$ is worthless with respect to the stock price
movement over a given time interval and for a given accuracy
$\theta$  in stock prices if, with probability $1,$ the path
followed by $S_t$ will not breach the curve (either from above or
from below) over that time interval.
\end{defn}

In other words, the problem we are interested in is that of finding
the minimum (maximum) value $S_{ml}\; (S_{mu})$ of the initial stock
price that will ensure that the lower (upper) curve  is worthless
over $[0,T].$  Let $P^{\, l}_t= \prob (S_t > B_l(t))$ be the
probability that $S_t > B_l(t).$ Then by Eq.~ \eqref{eq:gbm} we
clearly have

\begin{align}
\label{eq:pl0}  P^{\, l}_t &= \Phi(\omega^{\,l}_t) \notag \\[-4mm]
\intertext{ where }
                \omega^{\,l}_t &= \frac{(\mu- \sigma^2/2)t + \ln \left(S_0/ B_l(t)\right)}{\sigma
                \sqrt{t}},
\end{align}
and where $\Phi $ is the standard normal distribution function. It
is well known that $\Phi$ is a positive definite increasing function
satisfying $\lim_{x \ra \infty} \Phi(x)=1.$  However, this
convergence of $\Phi$ to $1$ is very fast and if we denote by $\pi=
\pi (\Phi)$ the accuracy of $\Phi,$ then it is easy to see that we
have for instance $\Phi (4.753424)=1,$ for $\pi= 10^{-6}.$ For a
given value of $\pi,$ let $\nu$ be the smallest number for which the
approximation $\Phi(\nu)=1$ holds, that is, such that $\phi(\nu)
\geq 1- \pi.$ We therefore have $P_t^l=1$ if and only if $\omega_t^l
\geq \nu,$ and solving this inequality for $S_0$ shows that
\begin{subequations}\label{eq:pl1}
\begin{align}
 P^{\, l}_t &=1 \Leftrightarrow   S_0 \geq S_l(t)\label{eq:pl1.1} \\[-4mm]
 \intertext{ where }
S_l (t)&= B_l (t)  e^{(\nu \sigma \sqrt{t}- \mu_1 t)} \label{eq:slt}
\\[-4mm]
\intertext{ and where }
\mu_1&= \mu - \sigma^2 /2 \label{eq:mu1} \\[-2mm] \notag
\end{align}
\end{subequations}
 Now, let $S_m$ be the maximum value of the continuous function $S_l(t)$ achieved
over the interval $I.$ Then by compactness we have $S_m= S_l(t_m)$
for some $t_m \in I.$ For $S_0 \geq S_m,$ Eq.~ \eqref{eq:pl1.1}
shows that $P^{\, l}_t=1$ for all $t \in I,$ that is the lower curve
is worthless. On the other hand, if $S_0 < S_m,$ then
\eqref{eq:pl1.1} shows again that $P^{\, l}_{t_m}\neq 1,$ and thus
the lower barrier isn't worthless. Consequently, $S_m$ is the
minimum value of the initial stock price above which the lower curve
becomes worthless, that is, $S_m= S_ml$ is the solution to Problem
1.\par

Similarly, let $P^u_t = \prob (S_t < B_u (t))$ be the probability
that $S_t < B_u (t).$ Then we have
\begin{align}
\label{eq:pu0}  P^u_t &= \Phi(\omega^u_t), \notag \\
\intertext{ where }
                \omega^u_t &= \frac{\ln\left(B^u(t)/S_0\right) - \mu_1 t}{\sigma
                \sqrt{t}},
\end{align}
and using again the fact that $P^u_t=1$ if and only if $\omega^u_t
\geq \nu,$ it readily follows that
\begin{subequations}\label{eq:pu1}
\begin{align}
 P^u_t &=1 \Leftrightarrow   S_0 \leq S_u(t),\label{eq:pu1.1} \\[-4mm]
 \intertext{ where }
S_u (t)&= B_u (t)  e^{-(\nu \sigma \sqrt{t}+ \mu_1 t)}.
\label{eq:sut}
\end{align}
\end{subequations}
Consequently, if we let $S_{\sc M}$ be the minimum value of the
continuous function $S_u(t)$ over the interval $I,$ then it can be
shown in a similar manner as in the preceding case for the lower
curve that $S_{\sc M}$ is the maximum value of the initial stock
price $S_0$ below which the upper curve is worthless. That is,
$S_{\sc M}= S_{mu}$ is the solution to Problem $2.$ We have thus
obtained the following result.
\begin{thm}\label{th:critic}
Suppose that the stock price $S_t$ follows a geometric Brownian
motion of the form \eqref{eq:gbm}. Suppose also that the two
functions $S_l (t)$ and $S_u(t)$  given by \eqref{eq:slt} and
\eqref{eq:sut}, respectively, are continuous and considered over the
time interval $I=[0,T].$ Let $S_{ml}$ be the maximum value of
$S_l(t)$ and $S_{mu}$ the minimum value of  $S_u(t).$ Then we have
\begin{enumerate}
\item[(a)] $S_{ml}$ is the minimum value of the initial stock price  $S_0$ above
which the curve $S= B_l(t)$ becomes worthless.
\item[(b)] $S_{mu}$ is the maximum value of the initial stock price
$S_0$ below which the curve $S= B_u (t)$ becomes worthless.
\end{enumerate}
\end{thm}

\section{Applications to barrier options}
\label{s:applications} We now consider more explicitly the problem
of determining when exactly a barrier option contingent on a stock
price $S_t$ will degenerate into a simpler type of option, as a
result of a change in the option's pricing parameters.  All barrier
options will be assumed to have, in general curved barriers, and the
underlying asset price $S_t$ is supposed to follow a geometric
Brownian motion of the form \eqref{eq:gbm}. We shall also denote by
$B_l(t)$ and by $B_u (t)$ the lower barrier and the upper barrier,
respectively, and if any, for a given barrier option.
\begin{defn}
We say that a barrier option of a given type is a typical barrier of
that type if none of its barriers can be considered worthless for
the corresponding parameter set. That is, if it cannot be priced as
a barrier option of a different type with the same parameter set.
\end{defn}

The equations \eqref{eq:slt} and \eqref{eq:sut} show that in
addition to the parameters of the curved barriers, the critical
values $S_{ml}$ and $S_{mu}$ of Theorem~\ref{th:critic} depend only
on the drift and volatility parameters $\mu$ and $\sigma$ of the
stock price $S_t,$ and on the expiry time $T.$ In a risk neutral
world, we have $\mu= r - \rho,$ where $r$ is the interest rate and
$\rho$ is the continuous dividend yield, if any, paid by the
underlying asset. However, the value of $\rho,$ is of no importance
for our analysis and we may assume without loss of generality that
$\rho=0.$ By the fixed parameter set for a given option, we shall
therefore refer to the other parameters $T, \sigma,$ and $r,$ as
well as the initial asset price $S_0$, the pay-off function
parameters such the strike price $K,$ and the curved barriers'
parameters.
\begin{thm} \label{th:dao}
Consider a down-and-out barrier option with  {\rm (}lower {\rm )}
barrier $S~=~B_l(t),$ and a given parameter set.
\begin{enumerate}
\item[(a)] If $S_0 \geq S_{ml},$ the barrier is worthless and the
option is equivalent to a vanilla option with the same parameter
set.
\item[(b)] The option is a typical down-and-out barrier option if and only
if \\$B_l(0)< S_0 < S_{ml}.$
\end{enumerate}
\end{thm}
\begin{proof}
Part (a) of the theorem readily follows from the definition of
$S_{ml},$ and the fact that a down-and-out option with a worthless
 barrier is naturally equivalent to a vanilla option with the same parameter
set. On the other hand, if $B_l(0)< S_0 < S_{ml},$ there is no
certainty that the barrier will not be breached before the expiry
date $T,$ and thus the option is necessarily a typical down-and-out
barrier option.
\end{proof}

\begin{thm} \label{th:uao}
Consider an up-and-out barrier option with {\rm (}upper{\rm )}
barrier $S~=~B_u(t),$ and a given parameter set.
\begin{enumerate}
\item[(a)] If $S_0 \leq S_{mu},$ the barrier is worthless and the
option is equivalent to a vanilla option with the same parameter
set.
\item[(b)] The option is a typical up-and-out barrier option if and only
if \\$S_{mu}< S_0 < B_u(0).$
\end{enumerate}
\end{thm}
\begin{proof}
The proof also follows from the definition of $S_{mu},$ and is
similar to that of Theorem~\ref{th:dao}.
\end{proof}
The following result is also straightforward from the definitions of
$S_{ml}$ and $S_{mu}$ given by Theorem~ \ref{th:critic}.
\begin{thm} \label{th:db}
Consider a double barrier option with lower barrier $S~=~B_l(t)$ and
upper barrier $S~=~B_u(t),$ and a fixed parameter set.
\begin{enumerate}
\topsep = 2.5mm
\itemsep = 1mm
\item[(a)] If $S_0 < S_{ml}, S_{mu},$ the option degenerates into a
down-and-out barrier option.
\item[(b)] If $S_{ml}, S_{mu} < S_0,$ the option degenerates into an
up-and-out barrier option.
\item[(c)] The option is equivalent to the vanilla option with the
           same parameter set if and only if  $S_{ml} \leq S_0 \leq S_{mu}.$
\item[(d)] The option is  a typical double barrier
option if and only if $S_{mu} < S_0 < S_{ml}.$
\end{enumerate}
\end{thm}
It should be noted that the results of the  theorems \ref{th:dao} -
\ref{th:db} do not depend on the pay-off function of the options
considered, and consequently they may be applied to all options
endowed with barriers, and in particular they apply to both barrier
call and barrier put options, European as well as American options,
and options including a rebate payment. The importance of these
theorems also stem from the fact that barriers can be added to
virtually all sorts of existing options, and this is one of the
reasons why barrier options have become so popular. An important
particular case is that of constant barriers. In such case, the
lower and upper curves, if any, for a given option can be denoted by
$S=B_l$ and $S=B_u,$ respectively, where $B_l$ and $B_u$ are
constants, and we can therefore give  more explicit expressions for
the critical values $S_{ml}$ and $S_{mu}.$ It is easy to see that
the  functions $S_l(t)$ and $S_u(t)$ given by the equations
\eqref{eq:slt} and \eqref{eq:sut} have in this particular case a
common expression for the turning point $t_p,$ when it exists,
given by
\begin{equation} \label{eq:tp}  t_p= \left(  \frac{\nu \sigma}{2 \mu_1 }\right)^2.
\end{equation}
Consequently we have the following explicit expressions for the
critical asset prices $S_{ml}$ and $S_{mu}.$
\begin{subequations}\label{eq:fcritic}
\begin{align}
 S_{ml} &= \begin{cases} B_{l \,} e^{(\nu \sigma
\sqrt{T} - \mu_1 T)},& \text{ if  $\mu_1 \leq 0$ or $ t_p \geq T$ }\\
B_{l \,}  e^{(\nu \sigma \sqrt{t_{p}} - \mu_1 t_{p})}, &\text{
otherwise}
\end{cases} \label{eq:smlf}\\
\intertext{ and }
S_{mu} &= \begin{cases} B_{u \,}e^{-(\nu \sigma \sqrt{T} + \mu_1 T
)}, & \text{ if  $ \mu_1 \geq 0 $
or $ t_{p} \geq T$ }\\
B_{u \,}e^{-(\nu \sigma \sqrt{t_{p}} + \mu_1 t_{p} )}, & \text{
otherwise.}
\end{cases} \label{eq:smuf}
\end{align}
\end{subequations}
Although to the best of our knowledge the critical value $S_{ml}$
and $S_{mu}$ are considered for curved barriers only for the first
time in this paper, it should be noted that equations of the form
\eqref{eq:fcritic} have been used in \cite{ndogbar1} for an optimal
determination of the solution domain in the numerical pricing of
barrier options with flat barriers. It's however, also the first
time that these critical values are applied to the classification of
barrier options.
\section{Numerical test}
\label{s:test}  The choice of the value of $\nu$ in the equations
\eqref{eq:slt} and \eqref{eq:sut} is clearly crucial. Indeed, these
equations show that the larger $\nu$ is, the farther both $S_{ml}$
and $S_{mu}$ will be with respect to their corresponding barriers,
and vice versa. This means that, as expected, larger values of $\nu$
yield critical values $S_{ml}$ and $S_{mu}$ which ensure with a
better certainty that the barriers are worthless, but which are
however not optimal. Inversely, smaller values of $\nu$ will yield
critical asset prices which are wrong by being too close to the
barrier, and this is just a consequence of the fact that the
equality $\Phi (\nu)=1$ is wrong for such values of $\nu$. Ideally,
$\nu$ should be as indicated the smallest number for which the
equality $\Phi (\nu)=1$ holds. However, for any numerical
determination, the constant $\nu,$ and consequently both $S_{ml}$
and $S_{mu},$ are very sensitive to the level of accuracy required
in option prices.\par
 A numerical value for $S_{ml}$ and $S_{mu}$ can be found by trial and
error on computing systems such as {\sc mathematica}. For example in
the case of a down-and-out barrier option, $S_{ml}$ is the smallest
value of the initial price $S_0$ for which the option price is the
same as that of the vanilla option with the same parameters. We can
then use such a numerical value to find the value of $\nu$ that
yields the same value of $S_{ml}$ given by Theorem~ \ref{th:critic}.
Next, using this choice of $\nu,$ and the same accuracy in option
prices, we can check whether  numerically computed critical asset
prices match those given by Theorem~ \ref{th:critic}, for different
sets of parameters.\par

\begin{table}[htb]
\parbox[]{4.1in}{\caption{\label{tb:test1} \protect \footnotesize Numerical test
of $S_{ml}$ for a down-and-out barrier option. The fixed parameter
set is $K=100, B_l=70,$ and  $r=0.10.$}}
\begin{tabular*}{3.1in}{ c c c c c c}
\hline \\[0.3mm]
T && $ \sigma$ & \parbox[c]{1 in}{ $\quad {\rm An} (S_{ml}$) \\
with
$\nu = 4.9 $}& $ {\rm N} (S_{ml})$ & $\nu$ \\[4mm] \hline \\[0.5mm]
0.25 &&  0.15   & 98.87      & 94.852       &4.347       \\
0.25 &&  0.30   &144.00      &156.744       &5.465       \\
0.50 &&  0.15   &112.60      &112.000       &4.850       \\
0.50 &&  0.30   &192.00      &229.999       &6.129       \\ \hline
\\[1.0mm]

\end{tabular*}
\parbox[t]{3.1in}{  \protect \footnotesize For $x=S_{ml},$  ${\rm An}(x)$
is the analytical value of $x$ given by Theorem \ref{th:critic}, and
$N(x)$ is the corresponding numerical value computed by trial and
error.}
\end{table}
%
First, we show by an example the effect of accuracy in  option
prices on the critical asset prices and on $\nu,$ by considering the
case of a down-and-out barrier option. We assume that the parameter
set is $T=1/4, K=100, B_l= 70, \sigma = 0.30$ and $r=0.10.$ Setting
$x= S_{ml},$ we see that when accuracy in option prices is to within
$2$ digits, we have $(x, \nu)=(77.182, 0.76),$ to within $6$ digits,
$(x, \nu)= (97.000, 2.267),$ and to within $17$ digits we have $(x,
\nu)=(156.744, 5.465).$ Intuitively, this clearly shows that  higher
accuracy yields  better values for both $\nu$ and the critical
values.\par

Table \ref{tb:test1} reports the test of $S_{ml}$ for a down-and-out
barrier option with a flat barrier and with fixed parameter set
$K=100, B_l =70, r=0.10$ and for variable values of the expiry date
$T$ and volatility parameters $\sigma.$ In the table,  ${\rm An}
(x)$ is the analytical value of $x=S_{ml}$ as given by Theorem
\ref{th:critic}, while $N(x)$ is the corresponding numerical value
computed by trial and error. The third column of the table contains
analytical values of $S_{ml}$ computed with a choice of $\nu=4.9$
according to the result of Theorem~ \ref{th:critic}, and more
specifically by using Eq.~ \eqref{eq:smlf}. Column $4$ gives the
actual numerical value of $S_{ml}$ for each parameter set, while the
last column gives the value of $\nu$ for which the analytical value
of  $S_{ml}$ given by Eq.~  \eqref{eq:smlf} would agree with the
numerical value given in the fourth column.\par
  The table shows that the value of $\nu=4.9,$ which is closely the
smallest number  $y$ for which {\sc mathematica} gives $\Phi (y)=1,$
is an appropriate value for Eq.~  \eqref{eq:slt} for relatively
small volatilities such as $\sigma=0.15.$ It also shows however that
$\nu$ should be larger for larger volatilities. Unfortunately, we
don't know how exactly the choice of $\nu$ changes with the
volatility, and with other pricing parameters. Nevertheless, Theorem
 \ref{th:critic} can be used efficiently when only approximate
 values of $S_{ml}$ and $S_{mu}$ are requested~ \cite{ndogbar1}. Thanks to
 the non dimensional nature of the parameter $\nu,$ the equations
 ~\eqref{eq:slt} and \eqref{eq:sut} together with Theorem~
 \ref{th:critic} provide all the properties of the critical assets
 prices and show how they depend in particular on the parameters $T,
 \sigma,$ and $r.$
\section{Concluding Remarks}\label{s:conclusion}
In this paper, we've found with very simple arguments in Theorem~
\ref{th:critic} critical stock prices that determine when exactly a
given stock price becomes worthless with respect to some given
curves, and we've given some applications of these critical values
to barrier options. This simple determination gives fully explicit
expressions for  the critical stock prices $S_{ml}$ and $S_{mu},$
although it remains approximate due to the presence of the
approximate parameter $\nu$ in these expressions. However, the
explicit nature of this determination, as expressed for example in
the particular case of flat barriers in  \eqref{eq:fcritic} ,
provide us with simple means to understand the properties of this
critical asset prices. Indeed, Table~ \ref{tb:test1} intuitively
confirms all the effects of the expiry date $T$ and volatility
$\sigma$ on these critical stock prices.\par
   Although the comparison in Table \ref{tb:test1} between actual
numerical values and corresponding semi-analytical values of the
critical prices displays some discrepancies, such values can be
safely and efficiently used in certain cases such as the optimal
truncation of the solution domain in a numerical scheme, and
especially in the case of the difficult problem of discrete barrier
options pricing. Indeed, in the case of constant barriers, an
application of the critical asset prices given by \eqref{eq:fcritic}
to the pricing of discrete barrier options  has given  in
\cite{ndogbar1}  results closest to the only semi-analytical one
available \cite{fusai}, when compared with similar results obtained
with five other numerical methods.\par

There are ways of finding exact numerical values for $S_{ml}$ and
$S_{mu}.$ For instance, the probability $Q(S, t, T)$ that the stock
price will breach a given barrier before time $T,$ given that it has
value $S$ at time $t$ can be formulated as a first time exit
problem, and it is well-known \cite{wilmott1} that $Q(S, t, T)$
satisfies the Kolmogorov Backward equation
\begin{equation}\label{eq:kolmo} \frac{\pd Q}{\pd t} +
B(S, t) \frac{\pd Q}{\pd S} + \frac{1}{2} A(S, t) \frac{\pd ^2
Q}{\pd S^2}=0, \end{equation}
with  $A(S,t)= \sigma S$ and $B(S,t)= \mu S,$ and with some boundary
conditions that are easy to determine. For instance we have $Q(S, T,
T)=0,$ and $Q(S, t, T)=1$ on the boundary curves.  The corresponding
boundary value problem can be solved explicitly, at least by
reducing Eq.~  \eqref{eq:kolmo} to a simpler equation. But after all
such lengthy calculations the corresponding expressions for $S_{ml}$
and $S_{mu}$ are not likely to be found explicitly, although their
exact numerical values can be found in this way. Equations of the
form \eqref{eq:fcritic} therefore have the advantage of providing in
a much quicker and simpler manner, in terms of the non-dimensional
parameter $\nu,$ all the information needed for the properties of
the critical values $S_{ml}$ and $S_{mu}.$\par

By replacing the two curves in Theorem \ref{th:critic} by the
straight line determined by the strike price $K$ of a given European
option, Theorem \ref{th:critic} can also be used to determine the
initial asset price range beyond which the option becomes worthless.
There are certainly several other applications of this theorem in
option pricing and finance.

%



\end{document}